\documentclass[twocolumn, 5p]{elsarticle}

\usepackage{graphicx}
\usepackage{xcolor}
\usepackage{color}
\usepackage{amsmath,amsfonts,amssymb}
\usepackage{latexsym}
\usepackage[utf8]{inputenc}

\title{Simple cosmological model with inflation and late times acceleration}

\author{Marek Szyd{\l}owski}
\ead{marek.szydlowski@uj.edu.pl}
\address{Astronomical Observatory, Jagiellonian University, Orla 171, 30-244 Krakow, Poland}
\address{Mark Kac Complex Systems Research Centre, Jagiellonian University, {\L}ojasiewicza 11, 30-348 Krak{\'o}w, Poland}
\author{Aleksander Stachowski}
\ead{aleksander.stachowski@doctoral.uj.edu.pl}
\address{Astronomical Observatory, Jagiellonian University, Orla 171, 30-244 Krakow, Poland}

\begin{document}

\begin{abstract}
In the framework of polynomial Palatini cosmology, we investigate a simple cosmological homogeneous and isotropic model with matter in the Einstein frame. We show that in this model during cosmic evolution, it appears the early inflation and the accelerating phase of the expansion for the late times. In this frame we obtain the Friedmann equation with matter and dark energy in the form of a scalar field with the potential whose form is determined in a covariant way by the Ricci scalar of the FRW metric. The energy density of matter and dark energy are also parametrized through the Ricci scalar. The early inflation is obtained only for an infinitesimally small fraction of energy density of matter. Between the matter and dark energy, there exists interaction because the dark energy is decaying. For characterization of inflation we calculate the slow roll parameters and the constant roll parameter in terms of the Ricci scalar. We have found a characteristic behaviour of the time dependence of density of dark energy on the cosmic time following the logistic-like curve which interpolates two almost constant value phases. From the required numbers of $N$-folds we have found a bound on model parameter.
\end{abstract}

\maketitle

\section{Introduction}

While current astronomical observations favour the standard cosmological model \cite{Ade:2015xua}, the $\Lambda$CDM model plays only role an effective theory of the Universe which offers rather the description of the current properties of the Universe than its explanations. The origin of properties of the current Universe we should look for in the very early Universe. In this context a very simple inflation model was proposed by Starobinsky in 1980 \cite{Starobinsky:1980te}. This model attracted attention of cosmologists because it can explain some troubles of the $\Lambda$CDM model in a very simple way. Moreover, this evolutional scenario is generic and emerged in cosmology in different contexts \cite{Ade:2015xua}. 
In this model, the inflationary scenario of the Universe is driven by the higher quadratic term in the action which takes the form $S=\int\sqrt{-g}\left(R+\frac{R^2}{6M^2}\right)d^4 x$.

This model \cite{Mukhanov:1981xt, Starobinsky:1983zz} predicts that slow roll parameters
$n_s=1-\frac{2}{N}$ and $r=\frac{12}{N^2}$ where $N=50\sim 60$ is the number of e-folds before the end of inflation, are in good agreement with Planck 2013 data. 

On the other hand, from the viewpoint of the complete quantum theory of gravity, higher order corrections $\alpha'=1/M_s^2$ to the Einstein-Hilbert action are always expected i.e. 
\begin{multline}
S=\int\sqrt{-g}(R+c_2 \alpha' R^2+\\
 \sum_{i=3}c_i\alpha'^{i-1}R^i+\text{ other higher derivative terms } )d^4 x,
\end{multline}
where $c_i$ are the dimensionless couplings.

The higher derivative terms in action may also originate from
the supergravity \cite{Farakos:2013cqa, Ferrara:2013kca}.

The problem of the inflation in polynomial $f(R)$ cosmology was investigated in the metric formalism in \cite{Huang:2013hsb}, where the spectral index and tensor-to-scalar ratio were calculated in the $f(R)$ inflation model.

In this paper we will phenomenologically investigate the inflation model with a polynomial form of the potential in the Palatini formalism in the Einstein frame \cite{Stachowski:2016zio, Szydlowski:2017uuy}. For simplicity we truncate Taylor series on term $R^3$.

The main aim of the paper is to investigate how rigid is the Starobinsky model of inflation and can be disturbed by switching higher order terms. Therefore, our study are motivated by the stability investigation. If the Starobinsky model is stable it is in some sense generic. The standard Starobinsky model of inflation is formulated in the background of metric formulation of $f(R)$ modified gravity. In this paper we formulate $f(R)$ theory in the Palatini formalism which gives us equation of motion in the form of the second order equation. The inflation similarly to the Starobinsky approach is obtained after transition to the Einstein frame. We obtain the form of the potential for scalar field in the covariant form directly parametrized by the Ricci scalar in the Palatini formulation.

We investigate how the shape of the potential changes under changing of the parameter which measures fraction of the higher order term in assumed $f(R)$ formula.

In modern cosmology, the Starobinsky model of inflation plays a crucial role \cite{Starobinsky:1980te}. This  model of the cosmic inflation is considered as a source of the inflaton field---higher curvature corrections with respect to the Ricci scalar $R$ in the Einstein-Hilbert action of gravity of the type $R^2$.

The Starobinsky model seems to be distinguished among different alternative models of the inflation as predicting a low value of the scalar-to-tensor ratio $r$,
namely, it predicts that $r\sim12/N^2$, where $N$ is the number of e-foldings during inflation \cite{Kehagias:2013mya}.

The Starobinsky model is also favoured by experimental results \cite{Planck:2013jfk, Ade:2015lrj, Ade:2015tva, Ade:2015xua, Adam:2015rua} which give an upper bound on $r$ around the value of $0.1$. What it is important from the observational point of view the Starobinsky model is the model with the highest Bayesian evidence \cite{Ade:2015lrj}.
It is characteristic that the other types of models which also fit the data are actually equivalent to the Starobinsky model during inflation \cite{Kehagias:2013mya}. 

From the methodological point of view it is important that the Starobinsky model can be embedded in different domains of fundamental physics. The situation is some sense similar to that which is happen in mathematics where an important theorem has many references to the important theorems in different areas of mathematics. One can distinguish embedding into the supergravity \cite{Cecotti:1987sa, Cecotti:1987qe} and embedding into the superstring theory \cite{Kounnas:2014gda, Blumenhagen:2015qda, Ellis:2015kqa, Alvarez-Gaume:2015rwa, Broy:2015zba}.

In our paper we consider a new embedding of Starobinsky model into cosmology of the Palatini gravity. It will be demonstrated the emergence of the inflation as an endogenous dynamical effect in the Palatini formulation of gravity applied to the FRW cosmology.

\section{Cosmological equations for the polynomial $f(\hat R)$ theory in the Palatini formalism in the Einstein frame}

In the Palatini formalism, the gravity action for $f(\hat{R})$ gravity has the following form
\begin{equation}
S=S_{\text{g}}+S_{\text{m}}=\frac{1}{2}\int \sqrt{-g}f(\hat{R}) d^4 x+S_{\text{m}},\label{action}
\end{equation}
where $\hat{R}$ is the generalized Ricci scalar \cite{Allemandi:2004wn,Olmo:2011uz}.

Let $f''(\hat R) \neq 0 $. In this case, the action (\ref{action}) has the equivalent form \cite{DeFelice:2010aj, Sotiriou:2008rp, Capozziello:2015wsa}
\begin{multline}\label{action1}
 S(g_{\mu\nu}, \Gamma^\lambda_{\rho\sigma}, \chi)=\frac{1}{2}\int\mathrm{d}^4x\sqrt{-g}\left(f^\prime(\chi)(\hat R-\chi) + f(\chi) \right) \\ + S_m(g_{\mu\nu},\psi),
\end{multline}
We introduce a scalar field $\Phi=f'(\chi)$, where $\chi=\hat R$. Then the action (\ref{action1}) is given by the following form
\begin{multline}\label{actionP}
 S(g_{\mu\nu}, \Gamma^\lambda_{\rho\sigma},\Phi)=\frac{1}{2}\int\mathrm{d}^4x\sqrt{-g}\left(\Phi \hat R - U(\Phi) \right) \\ + S_m(g_{\mu\nu},\psi),
\end{multline}
where the potential $U(\Phi)$ is defined as
\begin{equation}\label{PotentialP}
 U_f(\Phi)\equiv U(\Phi)=\chi(\Phi)\Phi-f(\chi(\Phi))
\end{equation}
with $\Phi = \frac{d f(\chi)}{d\chi}$ and $\hat R\equiv \chi = \frac{d U(\Phi)}{d\Phi}$.

The equations of motion are obtained after the Palatini variation of the action (\ref{actionP}) 
\begin{subequations}	
 \begin{gather}
	\label{EOM-P}
	\Phi\left( \hat R_{\mu\nu} - \frac{1}{2} g_{\mu\nu} \hat R \right) + \frac{1}{2} g_{\mu\nu} U(\Phi) - T_{\mu\nu} = 0,\\
	\label{EOM-connectP}
	\hat{\nabla}_\lambda(\sqrt{-g}\Phi g^{\mu\nu})=0,\\
	\label{EOM-scalar-field-P}
	  \hat R - U'(\Phi) =0.
	\end{gather}
\end{subequations}
From equation (\ref{EOM-connectP}) we get that a metric connection $\hat \Gamma$ is a new (conformally related) metric $\bar g_{\mu\nu}=\Phi g_{\mu\nu}$; thus $\hat R_{\mu\nu}=\bar R_{\mu\nu}, \bar R= \bar g^{\mu\nu}\bar R_{\mu\nu}=\Phi^{-1} \hat R$ and $\bar g_{\mu\nu}\bar R=\ g_{\mu\nu}\hat R$.
We can obtain from the $g$-trace of equation (\ref{EOM-P}) a new structural equation
\begin{equation}\label{struc2}
  2U(\Phi)-U'(\Phi)\Phi=T.
\end{equation}
Let $\bar U(\phi)=U(\phi)/\Phi^2$, $\bar T_{\mu\nu}=\Phi^{-1}T_{\mu\nu}$. Then equations (\ref{EOM-P}) and (\ref{EOM-scalar-field-P}) can be rewritten in the following form
	\begin{gather}
	\label{EOM-P1}
	 \bar R_{\mu\nu} - \frac{1}{2} \bar g_{\mu\nu} \bar R = \bar T_{\mu\nu}- \frac{1}{2} \bar g_{\mu\nu} \bar U(\Phi), \\
	\label{EOM-scalar-field-P1}
	  \Phi\bar R - (\Phi^2\,\bar U(\Phi))' =0,
	\end{gather}
and we get the following structural equation
\begin{equation}\label{EOM-P1c}
 \Phi\,\bar U'(\Phi) + \bar T = 0.
\end{equation}
In this case the action for the metric $\bar g_{\mu\nu }$ and the scalar field $\Phi$ has the following form in the Einstein frame
\begin{equation}\label{action2}
 S(\bar g_{\mu\nu},\Phi)=\frac{1}{2}\int\mathrm{d}^4x\sqrt{-\bar g}\left(\bar R- \bar U(\Phi) \right) + S_m(\Phi^{-1}\bar g_{\mu\nu},\psi)
\end{equation}
with a non-minimal coupling between $\Phi$ and $\bar g_{\mu\nu}$
\begin{equation}\label{em_2}
    \bar T^{\mu\nu} =
-\frac{2}{\sqrt{-\bar g}} \frac{\delta}{\delta \bar g_{\mu\nu}}S_m  = (\bar\rho+\bar p)\bar u^{\mu}\bar u^{\nu}+ \bar p\bar g^{\mu\nu}=\Phi^{-3}T^{\mu\nu}~,
\end{equation}
$\bar u^\mu=\Phi^{- \frac{1}{2}}u^\mu$, $\bar\rho=\Phi^{-2}\rho,\ \bar p=\Phi^{-2}p$, $\bar T_{\mu\nu}= \Phi^{-1}T_{\mu\nu}, \ \bar T= \Phi^{-2} T$ (see e.g. \cite{Capozziello:2015wsa, Dabrowski:2008kx}).

We take the metric $\bar g_{\mu\nu}$ in the standard form of the FRW metric
\begin{equation}\label{frwb}
d\bar s^2=-d\bar t^2+\bar a^2(\bar t)\left[dr^2+r^2(d\theta^2+\sin^2\theta d\phi^2)\right],
\end{equation}
where $d\bar t=\Phi(t)^{\frac{1}{2}}\, dt$ and a new scale factor $\bar a(\bar t)=\Phi(\bar t)^{\frac{1}{2}}a(\bar t) $. 	
The cosmological equations for the barotropic matter are given by
\begin{equation}\label{frwb2}
3\bar H^2= \bar \rho_\Phi + \bar\rho_\text{m}, \quad 6\frac{\ddot{\bar a}}{\bar a}=2\bar\rho_\Phi -\bar{\rho}_\text{m}(1+3w)
\end{equation}
where
\begin{equation}\label{frwb3}
\bar\rho_\Phi={\frac{1}{2}}\bar U(\Phi),\quad \bar{\rho}_{\text{m}}=\rho_0\bar a^{-3(1+w)}\Phi^{\frac{1}{2}(3w-1)}
\end{equation}
and
$w=\bar p_{\text{m}} / \bar\rho_{\text{m}}= p_{\text{m}} / \rho_{\text{m}}$. The conservation equations has the following form
\begin{equation}\label{frwb4}
\dot{\bar{\rho}}_{\text{m}}+3\bar H\bar{\rho}_{\text{m}}(1+w)=-\dot{\bar{\rho}}_\Phi.\end{equation}

In this paper, we consider the Palatini $f(\hat{R})=\sum_{i=1}^n \gamma_i \hat{R}^i$ model in the Einstein frame, where $\gamma_1=1$. In this case, the potential $\bar U$ is given by the following formula
\begin{equation}
\bar U(\hat R)=2\bar\rho_\Phi(\hat R)=\frac{\sum_{i=1}^n (i-1)\gamma_i \hat{R}^i}{\left(\sum_{i=1}^n i\gamma_i \hat{R}^{i-1}\right)^2}
\end{equation}
and the scalar field $\Phi$ has the following form
\begin{equation}
\Phi(\hat R)=\frac{df(\hat R)}{d\hat R}=\sum_{i=1}^n i\gamma_i \hat{R}^{i-1}.
\end{equation}

\section{Inflation in $f(\hat R)=\hat R+\gamma \hat R^2+\delta \hat R^3$ theory in the Palatini formalism in the Einstein frame}

Let $f(\hat R)=\hat R+\gamma \hat R^2+\delta \hat R^3$. For this case
\begin{equation}
\bar U(\hat R)=\frac{\hat R^2(\gamma+2\delta \hat R)}{\left(1+2\gamma \hat R+3\delta \hat R^2\right)^2}
\end{equation}
and
\begin{equation}
\Phi=1+2\gamma\hat R+3\delta \hat R^2.
\end{equation}
For this parametrization, we can obtain, from structural equation (\ref{EOM-P1c}), a parametrization of $\bar\rho_\text{m}$ with respect to $\hat R$
\begin{equation}
\bar\rho_\text{m}(\hat R)=\frac{\hat R-\delta \hat R^3}{\left(1+2\gamma \hat R+3\delta \hat R^2\right)^2}-4\Lambda.
\end{equation}
In consequence, the Friedmann equation is given by the following equation
\begin{equation}
 3\bar H^2=\bar\rho_\text{m}(\hat R)+\frac{\bar U(\hat R)}{2}+\Lambda=\frac{\hat R(2+\gamma \hat R)}{2\left(1+2\gamma \hat R+3\delta \hat R^2\right)^2}-3\Lambda.
\end{equation} 

In this model the inflation appears when matter $\bar\rho_\text{m}$ is negligible with comparison to $\bar\rho_\phi$.

In statistical analysis the slow roll parameters are helpful in the estimation of model parameter in the inflation period \cite{Ade:2015xua}. These parameters are defined as
\begin{equation}
\epsilon=-\frac{\dot H}{H^2}\text{ and }\eta=2\epsilon-\frac{\dot\epsilon}{2H\epsilon}.
\end{equation}
In our model the slow roll parameters have the following form in the case when $\delta=0$
\begin{equation}
\epsilon=\frac{3}{2}\frac{\hat R-4\Lambda(1+2\gamma \hat R)^2}{\hat R+\frac{\gamma}{2}\hat R^2-3\Lambda (1+2\gamma \hat R)^2},
\end{equation}
\begin{equation}
\eta=5+\frac{3}{2(\gamma\hat R-1)}+\frac{\hat R(1+2\gamma \hat R)}{6\Lambda(1+\gamma\hat R)^2-\hat R(2+\gamma \hat R)}.
\end{equation}

From the Planck observations, we know a limit at a 2-$\sigma$ level of the values of the scalar spectral index $n_\text{s}$ and the tensor-to-scalar ratio $r$ ($n_\text{s}=0.9667\pm 0.0040$ and $r<0.113$ \cite{Ade:2015xua}). The relation between the scalar spectral index and the tensor-to-scalar ratio and the slow roll parameters are the following
\begin{equation}
n_\text{s}-1=-6\epsilon+2\eta \quad \text{and} \quad r=16\epsilon.
\end{equation}

\begin{figure}[ht]
\centering
\includegraphics[scale=1]{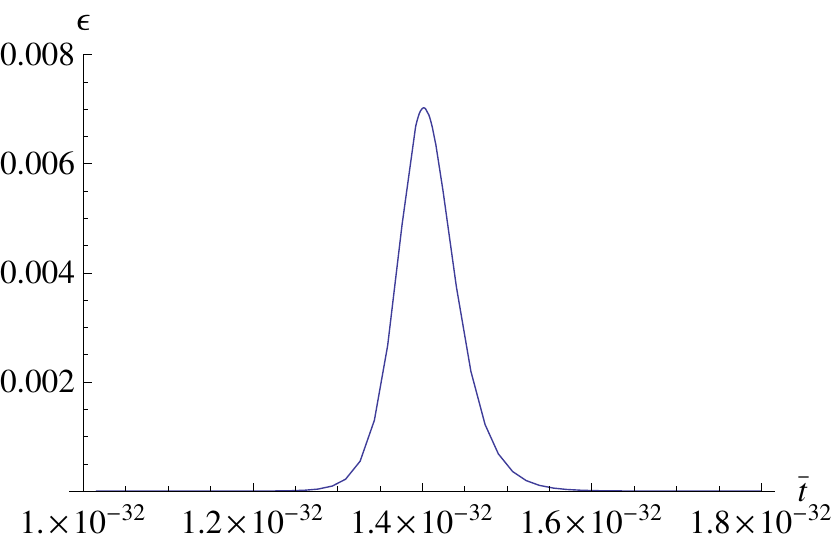}
\caption{Diagram presents the evolution of $\epsilon$ with respect to the cosmological time $\bar t$. The time is expressed in seconds. The value of the parameter $\gamma$ is assumed as $3.3\times 10^{-6}\frac{\text{s}^2 \text{Mpc}^2}{\text{km}^2}$. Note that $\epsilon$ is not a constant function when matter is not negligible (see Fig.~\ref{fig4}).}
\label{fig1}
\end{figure}

\begin{figure}[ht]
\centering
\includegraphics[scale=1]{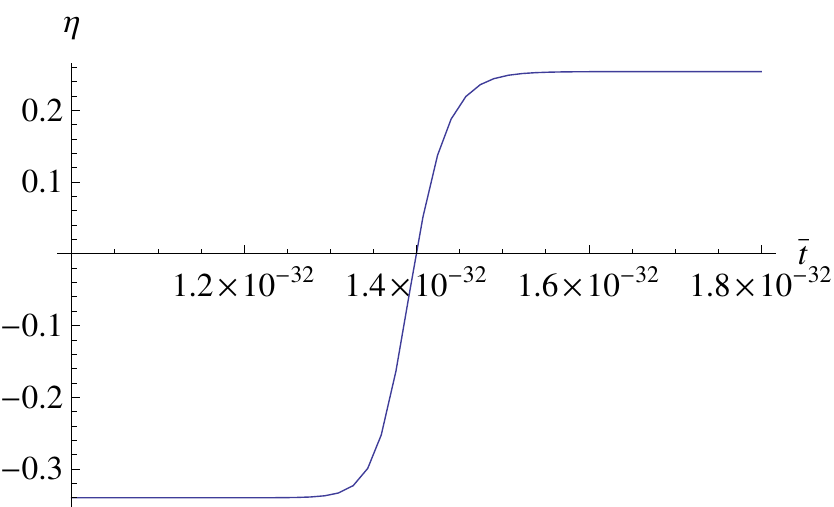}
\caption{Diagram presents the evolution of $\eta$ with respect to the cosmological time $\bar t$. The time is expressed in seconds. The value of the parameter $\gamma$ is assumed as $3.3\times 10^{-6}\frac{\text{s}^2 \text{Mpc}^2}{\text{km}^2}$. Note that $\eta$ is not a constant function when matter is not negligible (see Fig.~\ref{fig4}). It is interesting that the function $\eta$ is the logistic-like function type.}
\label{fig2}
\end{figure}

Because the slow roll parameters $\epsilon$ and $\eta$ cannot be treated as constant parameters in our model (see Figs \ref{fig1} and \ref{fig2}), then we cannot use these parameters to find the restriction on the parameter $\gamma$ from astronomical observations \cite{Ade:2015xua}.

For example, if we assume that $\frac{\Lambda}{3H_0^2}=0.6911$, where $H_0=67.74\frac{\text{km}}{\text{s Mpc}}$ \cite{Ade:2015xua} then we get that $3.285\times 10^{-6}\frac{\text{s Mpc}}{\text{km}}<\gamma<3.277\times 10^{-6}\frac{\text{s Mpc}}{\text{km}}$, $0<\Omega_{\text{m}} = \frac{\bar\rho_\text{m}}{3\bar H^2}<0.0047$ and $\Omega_{\Phi} = \frac{\bar\rho_\Phi}{3\bar H^2}\approx 0.50$. But this value of the parameter $\gamma$ is too large in order to explain the present evolution of the Universe. In consequence, the slow roll parameters are useless in the estimation of the parameter $\gamma$.

The slow roll parameter approximation is more restrictive than the constant roll condition \cite{Motohashi:2017aob, Motohashi:2017vdc}. The constant roll condition has the following form
\begin{equation}
\beta =\frac{\ddot\Phi}{\bar H\dot\Phi} = \text{const}.
\end{equation}
When $\beta\ll 1$ then we get the slow roll approximation.

\begin{figure}[ht]
\centering
\includegraphics[scale=1]{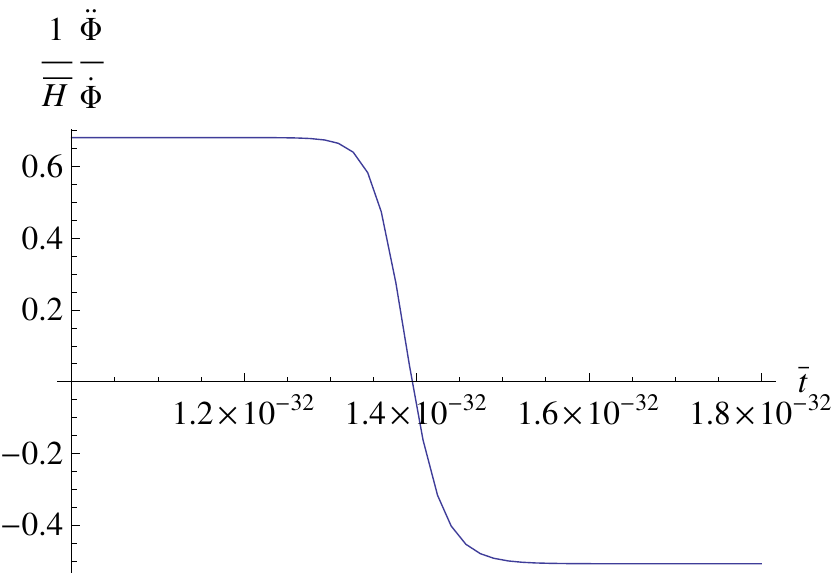}
\caption{Diagram presents the evolution of $\frac{\ddot\Phi}{H\dot\Phi}$ with respect to the cosmological time $\bar t$. The time is expressed in seconds. The value of the parameter $\gamma$ is assumed as $3.3\times 10^{-6}\frac{\text{s}^2 \text{Mpc}^2}{\text{km}^2}$. Note that $\frac{\ddot\Phi}{H\dot\Phi}$ is not a constant function when matter is not negligible (see Fig.~\ref{fig4}). It is interesting that $\frac{\ddot\Phi}{H\dot\Phi}$ function is the logistic-like function type.}
\label{fig3}
\end{figure}

In our case $\frac{\ddot\Phi}{\bar H\dot\Phi}$ is given by 
\begin{multline}
\frac{\ddot\Phi}{\bar H\dot\Phi}=4-240\gamma \Lambda+\frac{2}{1-24\gamma \Lambda}-192\gamma^2 \Lambda \hat R+\frac{9(36\gamma \Lambda-1)}{(\gamma\hat R-1)^2}+\\
\frac{12\Lambda+3(8\gamma\Lambda-1)\hat R}{(24\gamma\Lambda-1)\left(6\Lambda+\hat R(24\gamma\Lambda-2+\gamma(24\gamma\Lambda-1)\hat R)\right)}.
\end{multline}
when $\delta=0$. Because $\frac{\ddot\Phi}{\bar H\dot\Phi}$ is not constant (see Fig.~\ref{fig3}) at all the time but beyond the logistic-like type transition can be well approximated by a constant value. At this intermediate interval effects of matter do not become negligible. The constant roll inflation approximation is approximately valid beyond a short time during effects of matter stay very important (in consequence of the interaction between matter and dark energy).

\begin{figure}[ht]
\centering
\includegraphics[scale=1]{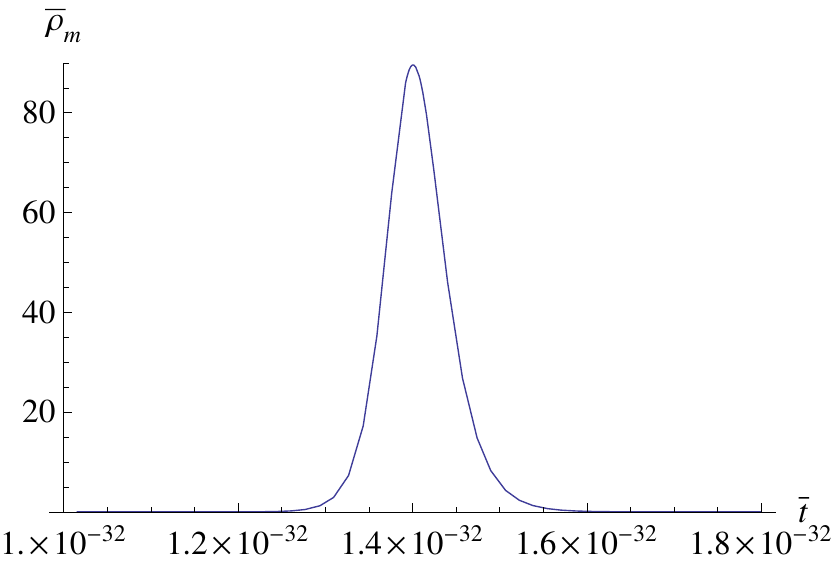}
\caption{Diagram presents the evolution of $\bar\rho_\text{m}$ with respect to the cosmological time $\bar t$. The time is expressed in seconds and $\bar\rho_\text{m}$ is expressed in $\frac{\text{km}^2}{\text{s}^2 \text{Mpc}^2}$. The value of $\gamma$ parameter is assumed as $3.3\times 10^{-6}\frac{\text{s}^2 \text{Mpc}^2}{\text{km}^2}$. Note that the maximum of this function is achieved when $\hat R=\frac{1}{2\gamma}$.}
\label{fig4}
\end{figure}

Fig.~\ref{fig1} presents the evolution of $\epsilon$ with respect to the cosmological time $\bar t$. We can see that $\epsilon$ is not a constant function when matter is not negligible (see Fig.~\ref{fig4}).

Fig.~\ref{fig2} demonstrates the evolution of $\eta$ with respect to the cosmological time $\bar t$. Note that $\eta$ is not a constant function when matter is not negligible (see Fig.~\ref{fig4}). The characteristic attribute of $\eta$ function is the shape of the logistic-like function.

Fig.~\ref{fig2} presents the evolution of $\frac{\ddot\Phi}{H\dot\Phi}$ with respect to the cosmological time $\bar t$. It is important that $\frac{\ddot\Phi}{H\dot\Phi}$ is not a constant function when matter is not negligible (see Fig.~\ref{fig4}). It is interesting that $\frac{\ddot\Phi}{H\dot\Phi}$ function is the logistic-like function type.

Note that $\beta=\frac{d\ln \dot\phi}{d\ln a}=\frac{\ddot\Phi}{H\dot\Phi}$ measures the elasticity of $\dot\phi$ with respect to the scale factor. When $\beta$ is constant then
\begin{equation}
\dot\phi \propto a^\beta.
\end{equation}
Therefore if $\beta$ is positive then $\dot\phi$ is a growing function of the scale factor. In the opposite case ($\beta<0$) $\dot\phi$ is an increasing function of the scale factor and goes to zero for the large value of the scale factor.

The slow roll approximation is achieved in our model when matter was negligible. Of course, the constant roll condition is respected automatically.

The evolution of matter in the inflation period can be divided into four phases. The first phase is when matter is negligible and the density of $\rho_\text{m}$ increases by the interaction with the potential $\rho_\Phi$. The second phase is when the matter cannot be negligible and its density still increases. In this phase the injection of matter is the most effective. After achieving of the maximum of the density of $\rho_\text{m}$ the third phase appears. In this phase matter still cannot be negligible but its density decreases. The last phase is when matter density decreases and is negligible.

The evolution of matter in the inflation period is presented in Fig.~\ref{fig4}. We can see all four phases of the evolution of matter. The maximum is achieved when
\begin{equation}
\hat R=\frac{1}{2\gamma}.
\end{equation}
In the maximum, the value of $\bar\rho_\text{m}$ is equal $\frac{1}{8 \gamma}-4\Lambda$.

\begin{figure}[ht]
\centering
\includegraphics[scale=1]{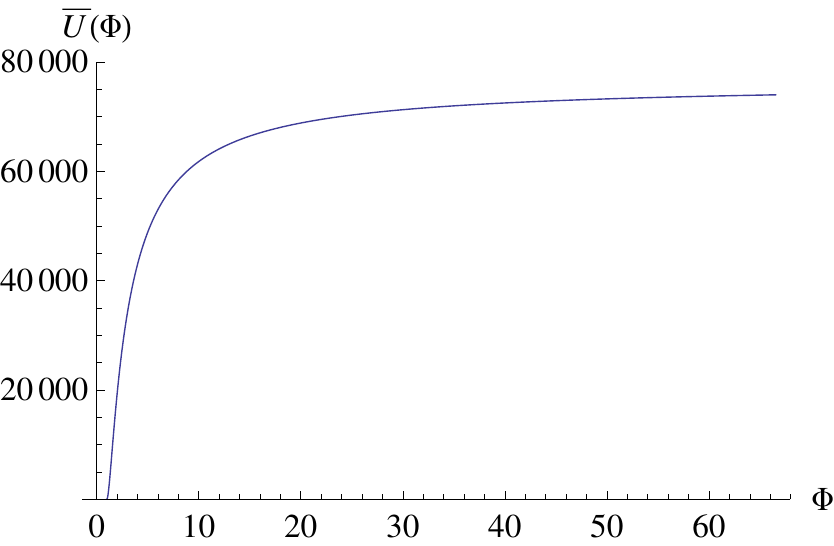}
\caption{Diagram presents typical behaviour of the function $\bar U(\Phi)$ for the case $f(\hat R)=\hat R+\gamma \hat R^2$.  The potential $\bar U(\Phi)$ is expressed in $\frac{\text{km}^2}{\text{s}^2 \text{Mpc}^2}$. Note that, for the large value of $\Phi$, function $\bar U(\Phi)$ has the plateau.}
\label{fig5}
\end{figure}

In details, the behaviour of the potential function $\bar U(\Phi)$ depends on the form of $f(\hat R)$. For the polynomial form of $f(\hat R)$, there are two cases. In the first case $f(\hat R)$ is in the form $f(\hat R)=\hat R+\gamma \hat R ^2$. The typical behaviour of the potential $\bar U(\Phi)$ for $f(\hat R)=\hat R+\gamma \hat R^2$ is presents in Fig.~\ref{fig5}. The characteristic attribute is a plateau for the large value of $\Phi$ like for the Starobinsky potential \cite{Starobinsky:1980te}. In this case the formula for the potential $\bar U(\Phi)$ has the following form
\begin{equation}
\bar U(\Phi)=\gamma \left(\frac{\Phi-1}{2\gamma \Phi}\right)^2.
\end{equation}

\begin{figure}[ht]
\centering
\includegraphics[scale=1]{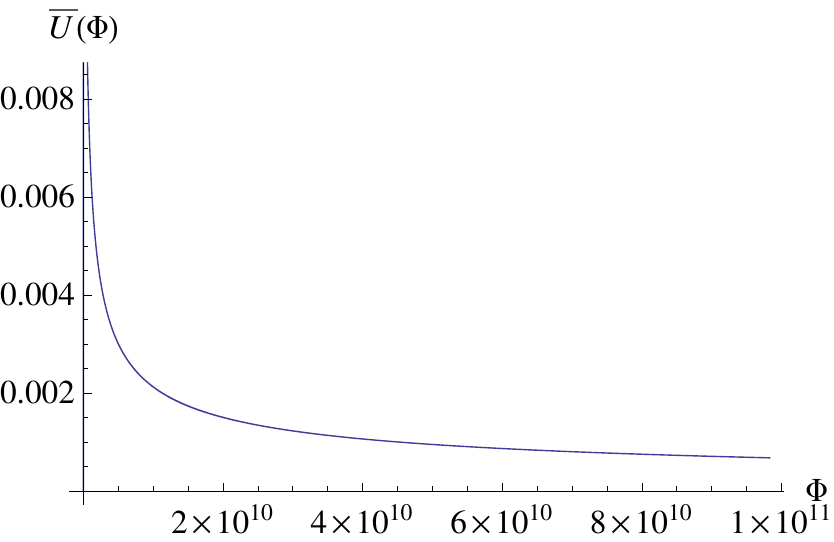}
\caption{Diagram presents typical behaviour of the function $\bar U(\Phi)$ for the case $f(\hat R)=\hat R+\gamma \hat R^2+\delta\hat R^2$. The potential $\bar U(\Phi)$ is expressed in $\frac{\text{km}^2}{\text{s}^2 \text{Mpc}^2}$. Note that, for the large value of $\Phi$, the function $\bar U(\Phi)$ decreases asymptotically to zero.}
\label{fig6}
\end{figure}

The second case is when $f(\hat R)$ is in the form $f(\hat R)=\hat R+\gamma \hat R ^2 +\sum_{i=2}^{n} \delta_i \hat R^{i+1}$. Then the potential $\bar U(\Phi)$ has not the plateau and decreases asymptotically to zero when $\Phi$ goes to the infinity. This situation is presented in Fig.~\ref{fig6}. The formula for the potential $\bar U(\Phi)$ for $f(\hat R)=\hat R+\gamma \hat R+\delta\hat R^2$ has the following form
\begin{equation}
\bar U(\Phi)=\frac{\left(\gamma-\sqrt{\gamma^2+3\delta(\Phi-1)}\right)^2 \left(\gamma+2\sqrt{\gamma^2+3\delta(\Phi-1)}\right)}{27 \delta^2\Phi^2}.
\end{equation}

In the context of inflation Ijjas et al. \cite{Ijjas:2013vea} pointed out the problem with the desired plateau in the behaviour of the potential of the scalar field. Such a choice seems to be unjustified because it requires that the power series expansion of potential $U$ with respect to $\Phi$ is cancelled at precise order in $\Phi$ to make the plateau appear.

In agreement with Ijjas et al. we obtain the plateau of the potential $\bar U(\Phi)$ only when $f(\hat R)=\hat R+\gamma \hat R^2$. For the higher order terms in the expansion of the $f(\hat{R})$, i.e. $R^3$ and higher, the potential monotically decreases to zero.

Now, we consider in details the inflation in both abovementioned cases with the potential expanded to the second order and third order with respect to $\Phi$.
In consequence, we study whether the plateau is necessary for appearing of the inflation in our model and whether inflation is possible for the model with a cut-off in higher order ($R^3$ and higher) expansion. 

\begin{figure}[ht]
\centering
\includegraphics[scale=1]{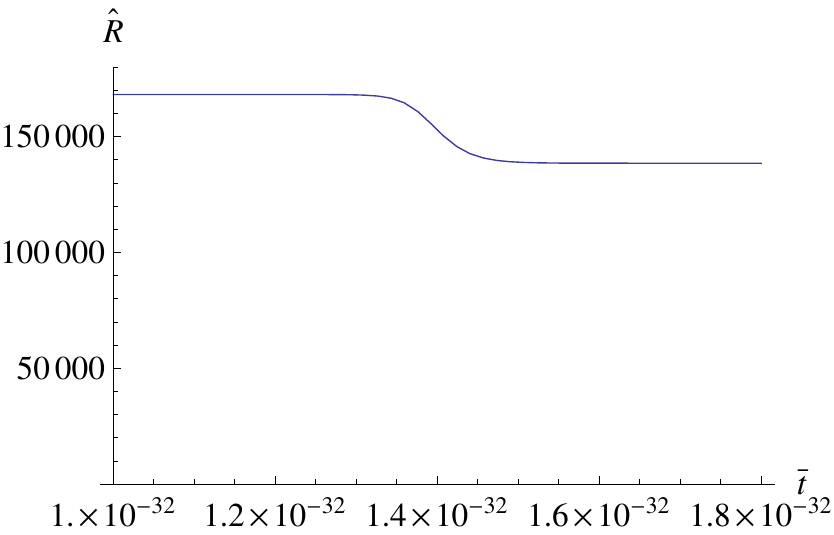}
\caption{Diagram presents the evolution of the Ricci scalar $\hat R$ with respect to the cosmological time $\bar t$. The time is expressed in seconds and the Ricci scalar $\hat R$ is expressed in $\frac{\text{km}^2}{\text{s}^2 \text{Mpc}^2}$. 
The transition phase is of logistic-like behaviour and is strictly correlated with a peak of matter density as it was shown in Fig.~4.}
\label{fig7}
\end{figure}

In the inflation period when the matter is negligible, the Ricci scalar $\hat R$ is constant. The evolution of the Ricci scalar $\hat R$ is presented in Fig.~\ref{fig7}. We can see three phases of the evolution of the Ricci scalar $\hat R$. The first phase is when matter is negligible and the density of $\rho_\text{m}$ is increased by an interaction with the potential $\rho_\Phi$. Then the Ricci scalar $\hat R$ is constant and is described by the following formula when $\delta=0$
\begin{equation}
\hat R=\frac{1-16\gamma \Lambda+\sqrt{1-32\gamma\Lambda}}{32\gamma^2 \Lambda}.
\end{equation}
The second phase is when the matter cannot be negligible. In this case, the Ricci scalar $\hat R$ decreases. The last phase is when matter density decreases and is negligible. Then the Ricci scalar $\hat R$ is constant and is equal
\begin{equation}
\hat R=\frac{1-16\gamma \Lambda-\sqrt{1-32\gamma\Lambda}}{32\gamma^2 \Lambda},
\end{equation}
when $\delta=0$. The function, which describes the evolution of the Ricci scalar $\hat R$ has the shape of the logistic-like function.

\begin{figure}[ht]
\centering
\includegraphics[scale=1]{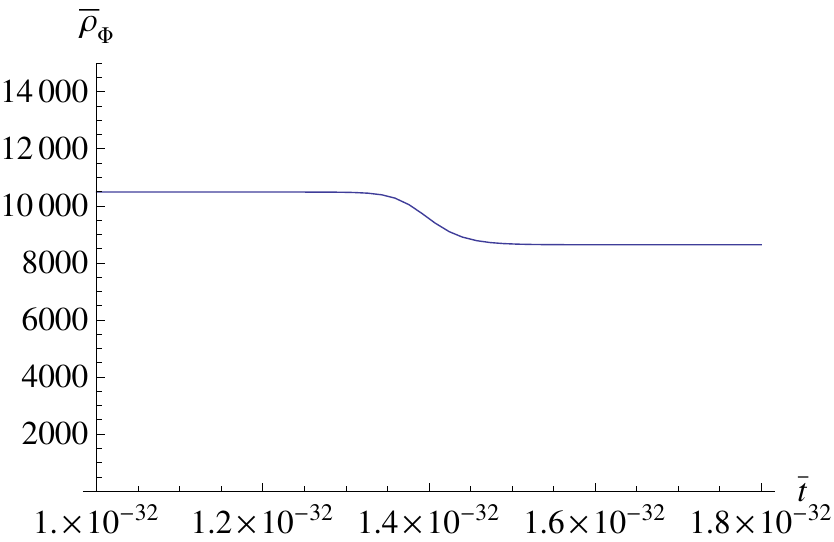}
\caption{Diagram presents the evolution of $\bar\rho_\Phi$ with respect to the cosmological time $\bar t$. The time is expressed in seconds and $\bar\rho_\Phi$ is expressed in $\frac{\text{km}^2}{\text{s}^2 \text{Mpc}^2}$. Note that $\bar\rho_\Phi$ is not a constant function when matter is not negligible (see Fig.~\ref{fig4}). It is interesting that the function $\bar\rho_\Phi$ is the logistic-like function type.}
\label{fig8}
\end{figure}

The evolution of $\rho_\Phi$, in the inflation period, similar qualitatively to the evolution of the Ricci scalar $\hat R$. We can find three phases. In the first phase, $\rho_\Phi$ is constant and is equal
\begin{equation}
\rho_\Phi=\frac{1-16\gamma \Lambda+\sqrt{1-32\gamma\Lambda}}{8\gamma}
\end{equation}
and in the last phase when $\rho_\Phi$ is also constant
\begin{equation}
\rho_\Phi=\frac{1-16\gamma \Lambda-\sqrt{1-32\gamma\Lambda}}{8\gamma}
\end{equation}
for $\delta=0$. 
The difference between $\rho_\Phi$ in the first and in the last phase is equal
\begin{equation}
\Delta\rho_\Phi=\frac{\sqrt{1-32\gamma\Lambda}}{4\gamma}\approx\frac{1}{4\gamma}.
\end{equation}
The evolution of $\rho_\Phi$ is presented in Fig.~\ref{fig8}. Our model predicts a phase of the early constant dark energy which is correlated with the inflation \cite{Pettorino:2013ia, Doran:2006kp}.

\begin{figure}[ht]
\centering
\includegraphics[scale=1]{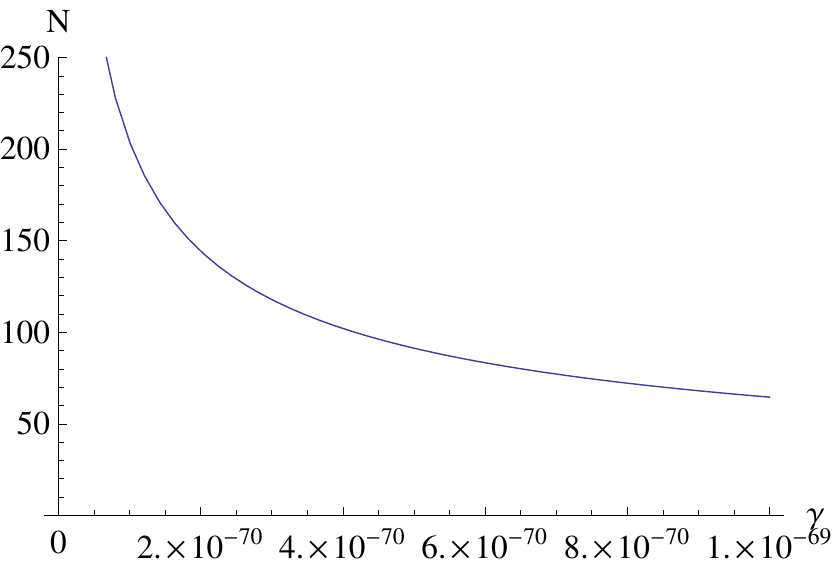}
\caption{Diagram presents the relation between the number of e-folds $N$ and the parameter $\gamma$. The parameter $\gamma$ is given in $\frac{\text{s}^2 \text{Mpc}^2}{\text{km}^2}$. We assume that $\delta=0$ and the inflation time is the order equal $10^{-32}\text{s}$ \cite{Ade:2014xna}.}
\label{fig9}
\end{figure}

\begin{figure}[ht]
\centering
\includegraphics[scale=1]{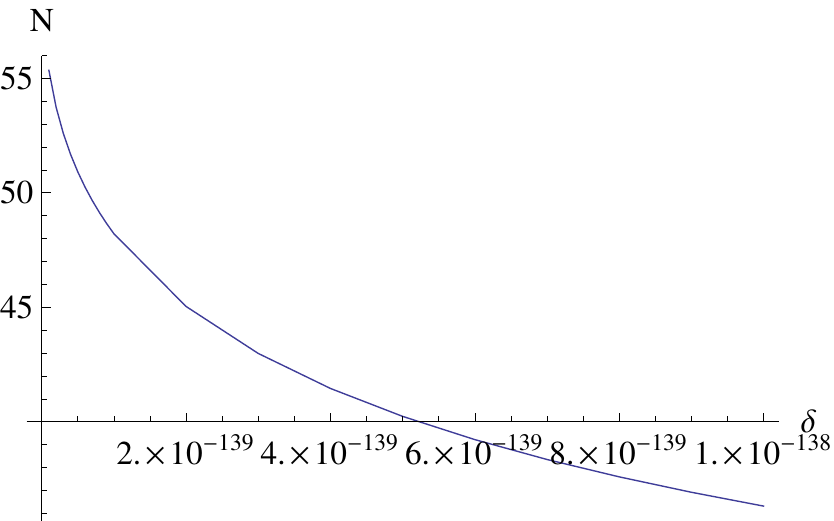}
\caption{Diagram presents the relation between the number of e-folds $N$ and the parameter $\delta$. The parameter $\delta$ is given in $\frac{\text{s}^4 \text{Mpc}^4}{\text{km}^4}$. We assume that $\gamma=1.16\times10^{-69}\frac{\text{s}^2 \text{Mpc}^2}{\text{km}^2}$ and the inflation time is the order equal $10^{-32}\text{s}$ \cite{Ade:2014xna}.}
\label{fig10}
\end{figure}

When $\delta=0$ number of e-folds in the first phase is equal
\begin{equation}
N=\frac{1}{4\sqrt{3}}\sqrt{\frac{1+\sqrt{1-32\gamma\Lambda}}{\gamma}}\left(\bar t_\text{fin}-\bar t_\text{ini}\right)\approx\frac{\bar t_\text{fin}-\bar t_\text{ini}}{4\sqrt{3\gamma}},
\end{equation}
where $\bar t_\text{fin}$ is time of the end of inflation and $\bar t_\text{ini}$ is time of beginning of the inflation. In the last phase
\begin{equation}
N=\frac{1}{4\sqrt{3}}\sqrt{\frac{1-\sqrt{1-32\gamma\Lambda}}{\gamma}}\left(\bar t_\text{fin}-\bar t_\text{ini}\right),
\end{equation}
Figs~\ref{fig9} and \ref{fig10} present number of e-folds in the first phase with respect to parameters $\gamma$ and $\delta$. In our model, the inflation appears only when $\delta\geq 0$.

If we assume that the parameter $\delta$ is equal zero and $N=50-60$ \cite{Cheng:2013iya} and the inflation time is the order equal $10^{-32}\text{s}$ \cite{Ade:2014xna} then the parameter $\gamma$ belongs to the interval $(1.16\times 10^{-69},\text{ }1.67\times 10^{-69})$. In consequence, the present value of $\frac{\rho_\Phi}{3 H^2}$ belongs to the interval $(3.41\times 10^{-61},\text{ }4.90\times 10^{-61})$. This means that the running dark energy is negligible in the present epoch and does not influence the acceleration of the present Universe. 

If the parameter $\delta \neq 0$ the number of e-folds is modified. For the parameter $\gamma$ belongs to the interval $(1.16\times 10^{-69},\text{ }1.67\times 10^{-69})$, we get the number of e-folds $N=50-60$, when the value of $\delta$ parameter belongs to the interval $(0,\text{ }6.4\times 10^{-140})$.

\section{Conclusions}

We are looking for cosmological model in which one can see of both the early inflation and the late times acceleration phase of expansion in the unique evolutional scenario. For this aim we study the cosmological model of polynomial $f(R)$ gravity cut on the $R^3$ term in the Palatini formalism in the Einstein frame. This model can be treated as an extension of the Starobinsky model which is formulated in the metric formalism. Our model is formulated in the Palatini formalism but it possesses analogous feature and its main advantage is simplicity. The model is reduced to the FRW model with matter and dark energy in the form of the homogeneous scalar field. Both energy densities of the matter as well as dark energy are determined by the Ricci scalar of the FRW metric. Therefore they are given in the covariant way. In the Einstein frame the energy density of the dark energy is fully determined by the potential of the scalar field. Because the density of dark energy is running, the interaction appears naturally between the matter and dark energy which can be also parametrized in the covariant way through the Ricci scalar. It is interesting that in our model it is possible to achieve some analytic formulas on the energy densities of dark energy and dark matter.

In our model, we have found that the plateau of the potential $\bar U(\Phi)$ is not necessary for appearing of the inflation \cite{Ijjas:2013vea}. In the expansion of function $f(R)$, the coefficient $\delta$ of the term $R^3$ affects the number of e-folds. The number of e-folds decreases for $\delta > 0$ with respect to the number of e-folds obtained for the model with the $f(R)$ expansion cut off at a quadratic term. In our model, the inflation appears only when $\delta\geq 0$. 

In the model if the matter is vanishing we obtain the eternal inflation following stationary solution $H=\text{const}$. This result is valid for the function $f(R)$ given by the polynomial form $f(\hat R)=\hat R+\gamma \hat R ^2 +\sum_{i=2}^{n} \delta_i \hat R^{i+1}$. Only for an infinitesimally small fraction of matter the inflation take places. The early inflation is studied in details in terms of slow roll parameters as well as using the conception of constant roll inflation. We calculate constant role parameter $\beta=\frac{d\ln \dot\phi}{d\ln a}$ which measure elasticity of $\dot\phi$ with respect to the scale factor. We have found the  characteristic type of the behaviour of the parameter $\beta$ following the logistic-like curve. One can distinguish four different phases in the time behaviour of the parameter $\beta$. In the first phase, effects of matter is negligible but due to an interaction with the dark energy sector, energy density of matter grows. During the second and third phase, effects of matter are not negligible. The later phase is characterized by vanishing effects of matter and the constant value of the Ricci scalar (and in consequence the constant value of energy density). During this phase dark energy dominates and the Universe behaves following the standard cosmological $\Lambda$CDM model.

Because the slow roll parameters are inadequate to constrain the model parameter we have found bound on model parameter $\gamma$ from the numbers of required $N$-folds. If we assume that $N=50-60$ \cite{Cheng:2013iya} then the parameter $\gamma$ belongs to the interval $(1.16\times 10^{-69},\text{ }1.67\times 10^{-69})$. For this interval of the parameter $\gamma$, we get the number of e-folds $N=50-60$, when the value of $\delta$ parameter belongs to the interval $(0,\text{ }6.4\times 10^{-140})$.

\section*{Acknowledgements}
We are very grateful of A. Borowiec and A. Krawiec for stimulating discussion and remarks.

\providecommand{\href}[2]{#2}\begingroup\raggedright\endgroup

\end{document}